\documentclass[aps,pra, 
 twocolumn
  ,showpacs,superscriptaddress]{revtex4-1}
\usepackage{graphicx}
\usepackage{amsmath}
\usepackage{amssymb}
\usepackage{amsfonts}
\usepackage{natbib}
\usepackage[breaklinks,colorlinks,
linkcolor=blue,citecolor=blue,urlcolor=blue]{hyperref}
\usepackage{dcolumn}
\newcolumntype{d}[1]{D{.}{.}{#1}}

\begin{document}
\title{
Path integral Monte Carlo determination of the fourth-order virial coefficient for unitary two-component 
  Fermi gas with zero-range interactions
}
\author{Yangqian Yan}
\affiliation{Department of Physics and Astronomy,
Washington State University,
  Pullman, Washington 99164-2814, USA}
\author{D. Blume}
\affiliation{Department of Physics and Astronomy,
Washington State University,
  Pullman, Washington 99164-2814, USA}
\date{\today}
\begin{abstract}
  The unitary equal-mass Fermi gas with zero-range interactions constitutes a
  paradigmatic model system that is relevant to atomic, condensed matter,
  nuclear, particle, and astro physics. This work determines the fourth-order
  virial coefficient $b_4$ of such a strongly-interacting Fermi gas using a
  customized \textit{ab initio} path integral Monte Carlo (PIMC) algorithm. In
  contrast to earlier theoretical results, which disagreed on the sign and
  magnitude of $b_4$, our $b_4$ agrees within error bars 
  with the experimentally determined
  value, thereby resolving an ongoing literature debate. Utilizing a trap
  regulator, our PIMC approach determines the fourth-order virial coefficient
  by directly sampling the partition function. An on-the-fly
  anti-symmetrization avoids the Thomas collapse and, combined with the
  use of the exact two-body zero-range propagator, establishes an efficient
  general means to treat small Fermi systems with zero-range interactions.
\end{abstract}
\maketitle
\textit{Introduction:}
Strongly-interacting Fermi gases manifest themselves 
in nature in different forms,
from neutrons in neutron stars~\cite{neutronstar95} to electrons in
solids~\cite{wen06}.
These systems are generally deemed difficult to treat theoretically
because of the lack of a small interaction parameter.
Superconductivity~\cite{BCS} and exotic states such as
fractional quantum hall~\cite{laughlin} or
Fulde-Ferrell-Larkin-Ovchinnikov~\cite{ff,larkin1964nonuniform,larkin1965inhomogeneous}
states have been observed or predicted to exist in these systems.
Ultracold Fermi gases~\cite{stringari08,blochreview}, 
which can nowadays be produced routinely in table-top experiments,
are ideal for studying strongly-interacting systems since
(i) the van der Waals interaction is short-ranged, which means that
it can be approximated by a contact potential that introduces 
a single length scale, i.e., the $s$-wave scattering length $a_s$;
and (ii) $a_s$ can be tuned at will utilizing Feshbach resonance techniques~\cite{eita10}.
When $a_s$ diverges, i.e., becomes infinitely large, the two-body contact potential does not define a length scale~\cite{heiselberg01}. 
Just like the non-interacting Fermi gas,
the properties of the unitary Fermi gas (Fermi gas with infinite $a_s$) 
are determined by two length scales,
the de Broglie wavelength $\lambda$ and interparticle spacing $\bar{r}$~\cite{ho04}.

At high temperature, $\lambda$ is much smaller than 
$\bar{r}$ and the grand canonical thermodynamic potential $\Omega$ can 
be expanded in terms of the fugacity~\cite{huang1987statistical,ho04l}.
The $n^{\text{th}}$-order expansion or virial coefficient $b_n$
is determined by the partition functions
of clusters containing $n$ or fewer fermions.
Since all thermodynamic properties at high temperature
can be derived from the virial coefficients $b_n$~\cite{liu13},
 the $b_n$'s are essential to understanding the normal state of 
 strongly-interacting Fermi gases.

While the second- and third-order virial coefficients are well understood~\cite{huang1987statistical,liu09,kaplan11,leyronas11,ensexp,liu13,b3bose,b3fermi},
none of the theoretical calculations for $b_4$~\cite{rakshit12,levinsen15,endo15,endoarXiv15} agree with the experimental data~\cite{ensexp,mitexp}.
 This letter rectifies this situation:  our theoretically determined $b_4$
 agrees with the experimentally determined value.
 Our approach uses a trap regulator~\cite{seminar89,trapregulator91} and employs the path integral Monte Carlo
 (PIMC) technique~\cite{ceperleyrev,boninsegni05}, with the contact interactions incorporated exactly via the
 two-body zero-range propagator~\cite{zr15}.
 The ``post-anti-symmetrization''~\cite{ceperleyrev,boninsegni05}, traditionally employed in PIMC calculations,
does not work
for the system with zero-range interactions, since the sampled paths shrink
due to the Thomas collapse, a well known phenomenon for bosons~\cite{thomas35,BraatenHammerReview}, to a single
point.
For bosons, the three-body Thomas collapse is cured by introducing an
additional scale or three-body parameter~\cite{BraatenHammerReview}. For fermions, such a three-body
parameter is not needed since the Pauli exclusion principle acts as an
effective three-body repulsion~\cite{petrov03,skor57}. Thus, rather than the standard
``post-anti-symmetrization'', we use an ``on-the-fly
scheme''~\cite{imada84,ceperleyfermion}, which
anti-symmetrizes at each imaginary time step. While the anti-symmetrization
is, within Monte Carlo frameworks, usually associated with 
the infamous Fermi sign
problem~\cite{FermiSignProblem,stratonovich1957method,ceperley1986quantum},
in our case it stabilizes the simulation and affords the use of significantly
smaller number of time slices than the use of finite-range interactions
would.
Our approach reproduces the trap regulated $b_3$ over a wide temperature
range.
We determine the trap regulated $b_4$ as a function of 
the temperature $T$.
In the low-temperature regime, we find agreement with Ref.~\cite{rakshit12}.
We separate the spin-balanced ($b_{2,2}/2$) and spin-imbalanced ($b_{3,1}$)
sub-cluster contributions to $b_4$,
$b_4=b_{3,1}+b_{2,2}/2$, and 
find $b_{2,2}<0$ and $b_{3,1}>0$ at all considered temperatures.
$b_{2,2}$ dominates at low $T$ and $b_{3,1}$ at high $T$.
Converting the trap regulated virial coefficient $b_4$ to that of the
homogeneous system 
using the local density approximation (LDA)~\cite{liu09}, 
we find agreement with the
experimentally determined values~\cite{ensexp,mitexp}.

\textit{Virial expansion framework:}
The $n^{\text{th}}$-order 
virial coefficient $b_n^{\text{hom}}$ of the homogeneous
system at unitarity is related to the high-temperature limit $b_n^{0}$ of the
harmonically trapped unitary system via 
$b_n^{\text{hom}}=n^{3/2} b_n^0$~\cite{trapregulator91,liu09}.
To determine $b_n^0$, we calculate the virial coefficient $b_n$ of the
harmonically trapped system for various temperatures
and then take the $T\to\infty$ limit.
The trap Hamiltonian $H(n_1,n_2)$ 
for $n_1$ particles of species $1$ and $n_2$ particles of species 
$2$ with interspecies $s$-wave interactions reads
\begin{align}
  \label{eq_Hamiltonian}
   H(n_1,n_2) = &\sum_{j=1}^{n_1+n_2} \left( \frac{-\hbar^2}{2 m} 
  \mathbf{\nabla}_{\mathbf{r}_j}^2 + 
  \frac{1}{2} m \omega^2 \mathbf{r}_j^2 \right) \nonumber\\
  & 
  + \sum_{i=1}^{n_1} \sum_{j=n_1+1}^{n_1+n_2}
  V_{2b}(\mathbf{r}_{i}-\mathbf{r}_{j}),
\end{align}
where $m$ denotes the mass of the particles, $\mathbf{r}_j$ the position
vector of the $j^{\text{th}}$ particle, $\omega$ the angular trapping
frequency, and  $V_{2b}$ the regularized Fermi-Huang
pseudopotential with infinite $a_s$~\cite{yang57}.
The grand canonical thermodynamic potential $\Omega$ can be written in terms
of the fugacities $z_i$ of species $i$,
\begin{equation}
  \Omega=-k_B T \ln\left(\sum_{n_1=0}^\infty\sum_{n_2=0}^\infty Q_{n_1,n_2} z_1^{n_1} z_2^{n_2}\right),
  \label{thermodynamicPotential}
\end{equation}
where $z_i$ is equal to $\exp[\mu_i/(k_B T)]$, 
$\mu_i$ is the chemical potential of
species $i$, and $Q_{n_1,n_2}$ is the canonical partition function for
$H(n_1,n_2)$,
\begin{align}
  \label{eq_canonicalPartFcta}
  Q_{n_1,n_2} = \; \mbox{Tr} \; \exp{[-H(n_1,n_2)/(k_B T)]}.
\end{align}
Here, Tr is the trace operator.
Defining $\Delta \Omega=\Omega-\Omega^{\text{ni}}$, where $\Omega^{\text{ni}}$ is the grand
canonical potential of the non-interacting system, and Taylor-expanding
around $z_1=z_2=0$~\cite{liu10,kevin12}, one finds
\begin{eqnarray}
  \label{eq_domega2}
  \Delta \Omega= - k_B T Q_{1,0} 
  \left( \sum_{n_1=1}^{\infty}\sum_{n_2=1}^{\infty} 
  b_{n_1,n_2} z_1^{n_1}  z_2^{n_2}
  \right).
  \label{deltathermodynamicPotential}
\end{eqnarray}
For spin-balanced systems, $z_1$ and $z_2$ are equal and
Eq.~(\ref{deltathermodynamicPotential}) reduces to
\begin{eqnarray}
  \label{eq_domega2simple}
  \Delta \Omega = -2 k_B T  Q_{1,0} \left(
  \sum_{n=2}^{\infty} b_n z^n \right),
\end{eqnarray}
where $b_2=b_{1,1}/2$, $b_{3}=(b_{1,2}+b_{2,1})/2$, and
$b_4=(b_{1,3}+b_{3,1}+b_{2,2})/2$ (note, one has $b_{2,1}=b_{1,2}$ and
$b_{3,1}=b_{1,3}$).
It is convenient to write 
the virial coefficients $b_{n_1,n_2}$ as 
\begin{eqnarray}
  b_{n_1,n_2} =\Delta b_{n_1,n_2} + b_{n_1,n_2}^{\rm{ref}},
\end{eqnarray}
where $b_{n_1,n_2}^{\rm{ref}}$  is determined by 
the virial coefficients $b_{j_1,j_2}$ and the canonical
partition functions $Q_{j_1,j_2}$ with $j_1+j_2 < n_1+n_2$.
The term $\Delta b_{n_1,n_2}=(Q_{n_1,n_2}-Q_{n_1,n_2}^{\text{ni}})/Q_{1,0}$, where $Q_{n_1,n_2}^{\text{ni}}=Q_{n_1,0}Q_{0,n_2}$,
in contrast, accounts for the ``new''
physics introduced by the interacting $(n_1,n_2)$ clusters~\cite{b4refnote}.

\textit{Contradicting literature results for $b_4$:}
The literature results are summarized in Table~\ref{tab1} (see also the
supplemental material~\cite{supnote}).
\begin{table}[!pb]
  \centering
  \caption{
    Summary of literature and PIMC results.
}
\begin{tabular}
  {d{2.7} d{2.10} c c}
    \hline
    \hline
    \multicolumn{1}{c}{$b_4^{\text{hom}}$}  &\multicolumn{1}{c}{$b_4^0$} & Ref. & comment\\
    0.096(15)&  0.01200(188) & \cite{ensexp} & ENS experiment\\
    0.096(10)&  0.01203(125) & \cite{mitexp} & MIT experiment\\
    -0.016(4) & -0.0020(5) & \cite{rakshit12} & sum-over-states approach\\
    0.06 & 0.0075 &\cite{levinsen15} & diagrammatic approach\\
    -0.063(1) & -0.007875(125) &\cite{endo15}& 3-body inspired
     conjecture\\
     0.078(18) & 0.0098(23) & & PIMC, this work\\
    \hline
    \hline
  \end{tabular}
  \label{tab1}
\end{table}
Two independent experiments find consistent values for the fourth-order virial
coefficient. The theoretical literature results, however,
disagree with these experimental results, reflecting the fact that the
fourth-order problem is highly non-trivial analytically and numerically.
Using a sum-over-states approach with an energy cutoff, Ref.~\cite{rakshit12}
obtained the low-temperature behavior of $b_4$. Assuming a monotonic
temperature dependence and extrapolating to the $T\to\infty$ limit,
Ref.~\cite{rakshit12} obtained $b_4^0$. It was concluded that more four-body
energies would need to be calculated explicitly to obtain $b_4$ reliably at
high temperature.
The fourth-order virial coefficient has also been obtained by a diagrammatic
approach, which included only a subset of the four-body free-space
diagrams~\cite{levinsen15}, and by applying a conjecture inspired by
three-body results~\cite{endo15,endoarXiv15}.

\textit{Customized PIMC algorithm:}
$\Delta b_{n_1,n_2}$ is determined by the 
partition function $Q_{n_1,n_2}$ of the interacting $(n_1,n_2)$ system ($Q_{n_1,n_2}$ is not known in general) and the partition function $Q_{n_1,n_2}^{\text{ni}}$ of the non-interacting $(n_1,n_2)$ system
($Q_{n_1,n_2}^{\text{ni}}$ is known analytically).
We calculate the ratio of the partition functions 
$Q_{n_1,n_2}^{\text{ni}}/Q_{n_1,n_2}$ using the PIMC technique.
Specifically, the simulation generates configurations according
to $Q_{n_1,n_2}$ and accumulates the ratio 
$Q_{n_1,n_2}^{\text{ni}}/Q_{n_1,n_2}$ as a
weight.
The reason for using the partition function of the unitary Fermi gas and
not that of the non-interacting gas as 
the ``guiding function''
is the following.
The probability density to find two unlike particles with vanishing interparticle
spacing is finite at unitarity and zero in the non-interacting limit.
If we used $Q_{n_1,n_2}^{\text{ni}}$ as the guiding function,
configurations in which two unlike particles 
are at the same spatial position would
be absent and the standard deviation of 
$Q_{n_1,n_2}/Q_{n_1,n_2}^{\text{ni}}$
would be infinite, rendering the expectation value
meaningless~\cite{infinitestdnote}.
  
  In the PIMC formulation, the partition function
  $Q_{n_1,n_2}^{\text{boltz}}(\beta)$ for Boltzmann particles (no exchange
  symmetries) at inverse temperature $\beta$, $\beta=(k_B T)^{-1}$, is written
  in terms of a product over density matrices at imaginary time 
  $\tau$,
\begin{equation}
  Q_{n_1,n_2}^{\text{boltz}}(\beta)=\int\dots\int \prod_{i=1}^{N}\rho(\mathbf{R}_i,\mathbf{R}_{i+1};\tau) d\mathbf{R}_1\dots d\mathbf{R}_N,
  \label{bosonpartition}
\end{equation}
where 
$\mathbf{R}_i$ collectively denotes the particle configurations at time
slice $i$, $\mathbf{R}_N=\mathbf{R}_1$, and $N=\beta/\tau$. 
For the two-component Fermi gas, the standard PIMC approach writes the partition
function as $Q_{n_1,n_2}=\mathcal{A}Q_{n_1,n_2}^{\text{boltz}}$, where
$\mathcal{A}$ is the anti-symmetrizer~\cite{antisym,ceperleyrev}. 
For the two-component Bose gas, the anti-symmetrizer
$\mathcal{A}$ is replaced by the symmetrizer $\mathcal{S}$.
$\mathcal{A}$ and $\mathcal{S}$ contain the same number and types of
terms; however, while all terms in $\mathcal{S}$ enter with a plus sign,
$\mathcal{A}$ contains alternating plus and minus signs.
Since the symmetrizer and anti-symmetrizer are, in the standard PIMC approach, 
evaluated stochastically,
the two-component Fermi and 
Bose gases are simulated using the same paths.
Expectation values, however, are accumulated with plus and minus signs 
for fermions and with plus signs only for bosons.
We refer to this standard approach as
post-symmetrization.
The bosonic system with interspecies two-body zero-range interactions but
without a three-body regulator would collapse to a single point; this is the
well-known Thomas collapse~\cite{thomas35}. Correspondingly, the fermionic
paths would also collapse, rendering the simulation meaningless. To get around
this problem, we developed a customized on-the-fly anti-symmetrization scheme,
which explicitly anti-symmetrizes the density matrix at each imaginary time
step,
\begin{equation}
  Q_{n_1,n_2}(\beta)=\int\dots\int \prod_{i=1}^{N}\mathcal{A}\rho(\mathbf{R}_i,\mathbf{R}_{i+1};\tau) d\mathbf{R}_1\dots d\mathbf{R}_N.
  \label{fermionpartition}
\end{equation}
The observable is then calculated using
\begin{equation}
  \frac{Q_{n_1,n_2}^{\text{ni}}}{Q_{n_1,n_2}}=\left<\prod_{i=1}^{N}\frac{\mathcal{A}\rho^{\text{ni}}(\mathbf{R}_i,\mathbf{R}_{i+1};\tau)}{\mathcal{A}\rho(\mathbf{R}_i,\mathbf{R}_{i+1};\tau)}\right>,
  \label{}
\end{equation}
where $\rho^{\text{ni}}$ denotes the density matrix 
for the non-interacting system and
$\left<\dots\right>$ the thermal average using paths generated for
the unitary Fermi gas using the
on-the-fly anti-symmetrization scheme.
Our simulation uses 
the pair-product
approximation~\cite{ceperleyrev,supnote} with
the exact two-body density matrix for zero-range
interactions. The on-the-fly scheme employed here is related to earlier
works~\cite{imada84,chin15}, which anti-symmetrized, as we do, at each
time slice. The key difference is that we employ a density matrix that
accounts for the interactions while the earlier works employed the
non-interacting density matrix together with the Trotter (or improved Trotter)
formula.

The on-the-fly anti-symmetrization scheme treats the $n_1!n_2!$ permutations
explicitly at each time slice, eliminating the need of the standard
stochastic ``permute move''.
As a consequence, the scheme is computationally prohibitively demanding for
large systems. For small systems, however, it is quite efficient for three
reasons: (i) The number of permutations is manageable for small $n_1+n_2$. 
(ii) The use of the zero-range interactions eliminates the need to
perform calculations for several different ranges of the underlying two-body
potential.
(iii) Compared to finite-range interactions~\cite{statistics14}, 
the number of time
slices needed to reach convergence for the zero-range interacting systems
considered here is rather small; e.g., our scheme yields
$Q_{3,1}^{\text{ni}}/Q_{3,1}$ at $E_{\text{ho}}/(k_B T)=0.8$  with $0.1\%$ error
using only $N=9$ imaginary time slices (here, $E_{\text{ho}}=\hbar\omega$). Within our approach, the key challenge
in determining $b_4$ reliably  at high temperature comes from the fact that
$\Delta b_{2,2}$, $\Delta b_{3,1}$, $b_{2,2}^{\text{ref}}$, and
$b_{3,1}^{\text{ref}}$ diverge, to leading order, as $(k_BT/E_{\text{ho}})^6$. This
implies that $b_{2,2}$ and $b_{3,1}$ are,
at high temperature, obtained by adding two
numbers of opposite sign and nearly equal magnitude. Thus, to obtain reliable
values at high temperature,
we need to determine our observables with high accuracy.
In practice, our available computer time limits us to $k_B T\le 2E_{\text{ho}}$ for
the (2,2) and (3,1) systems.

\begin{figure}
\includegraphics[angle=0,width=0.4\textwidth]{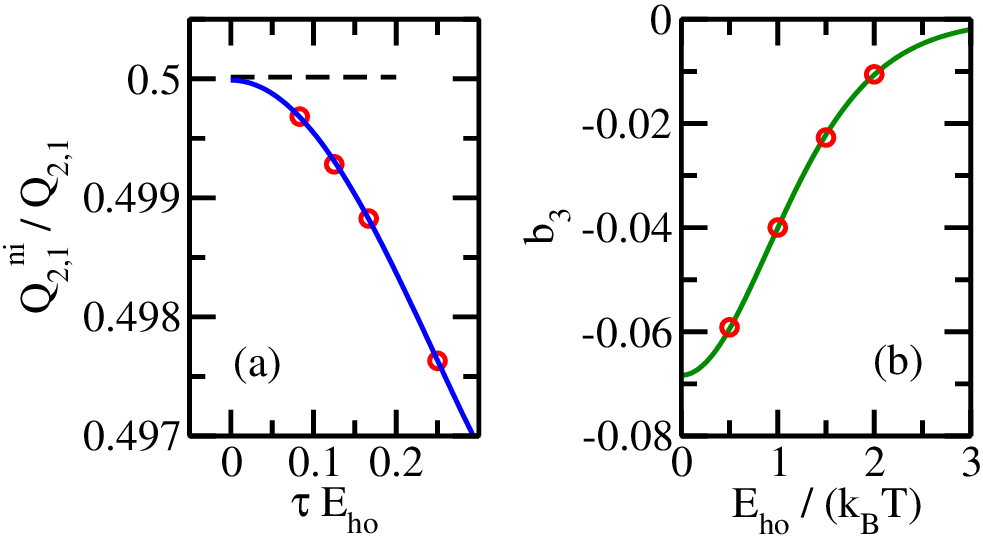}
\caption{(Color online)
  Benchmarking our PIMC results (circles) for the (2,1) system at unitarity
  through comparison with sum-over-states results.
  (a) The observable $Q_{2,1}^{\text{ni}}/Q_{2,1}$
  as a function of the 
  imaginary time step $\tau$ at $k_BT=E_{\text{ho}}$.
  The circles show our PIMC results.
  The error bars (not shown) are smaller than the symbol size.
  The solid line shows the fourth-order polynomial fit of the form $a+b
  \tau^2+c\tau^4$.
  The dashed line shows the sum-over-states results.
  (b) $b_3$ 
  as a function of $1/(k_B T)$.
  The circles show our PIMC results while the solid line shows 
  the sum-over-states results.
 }\label{figet1}
\end{figure}

\textit{PIMC results:}
To benchmark our customized PIMC algorithm, we apply it to the (2,1) system at
unitarity, for which $Q_{2,1}^{\text{ni}}/Q_{2,1}$ and $b_3$ can be calculated
with high accuracy for all temperatures using the sum-over-states approach~\cite{liu09}.
As an example, circles in Fig.~\ref{figet1}(a) show the quantity
$Q_{2,1}^{\text{ni}}/Q_{2,1}$ for $k_BT=E_{\text{ho}}$, obtained using our PIMC
algorithm, as a function of the imaginary time step $\tau$. The $\tau$
considered correspond to between $N=4$ and $10$ time slices.
The simulation is exact in the $\tau\to 0$ (or equivalently, $N\to \infty$) 
limit.
To extrapolate to the $\tau\to 0$ limit, we fit a fourth-order polynomial of
the form $a+b\tau^2+c\tau^4$ to the PIMC data [solid line in
Fig.~\ref{figet1}(a)]. Our extrapolated result of $0.499989(26)$ agrees within
error bars with the value of $0.500014$ [dashed line in Fig.~\ref{figet1}(a)]
obtained by the sum-over-states approach. Using the extrapolated $\tau\to 0$
values for $Q_{2,1}^{\text{ni}}/Q_{2,1}$ at various temperatures 
$T$, we obtain $b_3$ as
a function of $T$ [circles in Fig.~\ref{figet1}(b)]. The agreement with the
sum-over-states results [solid line in Fig.~\ref{figet1}(b)] is excellent for
all $T$ considered, demonstrating the reliability and accuracy of our PIMC
approach.

\begin{figure}[!tbp]
\centering
\includegraphics[angle=0,width=0.4\textwidth]{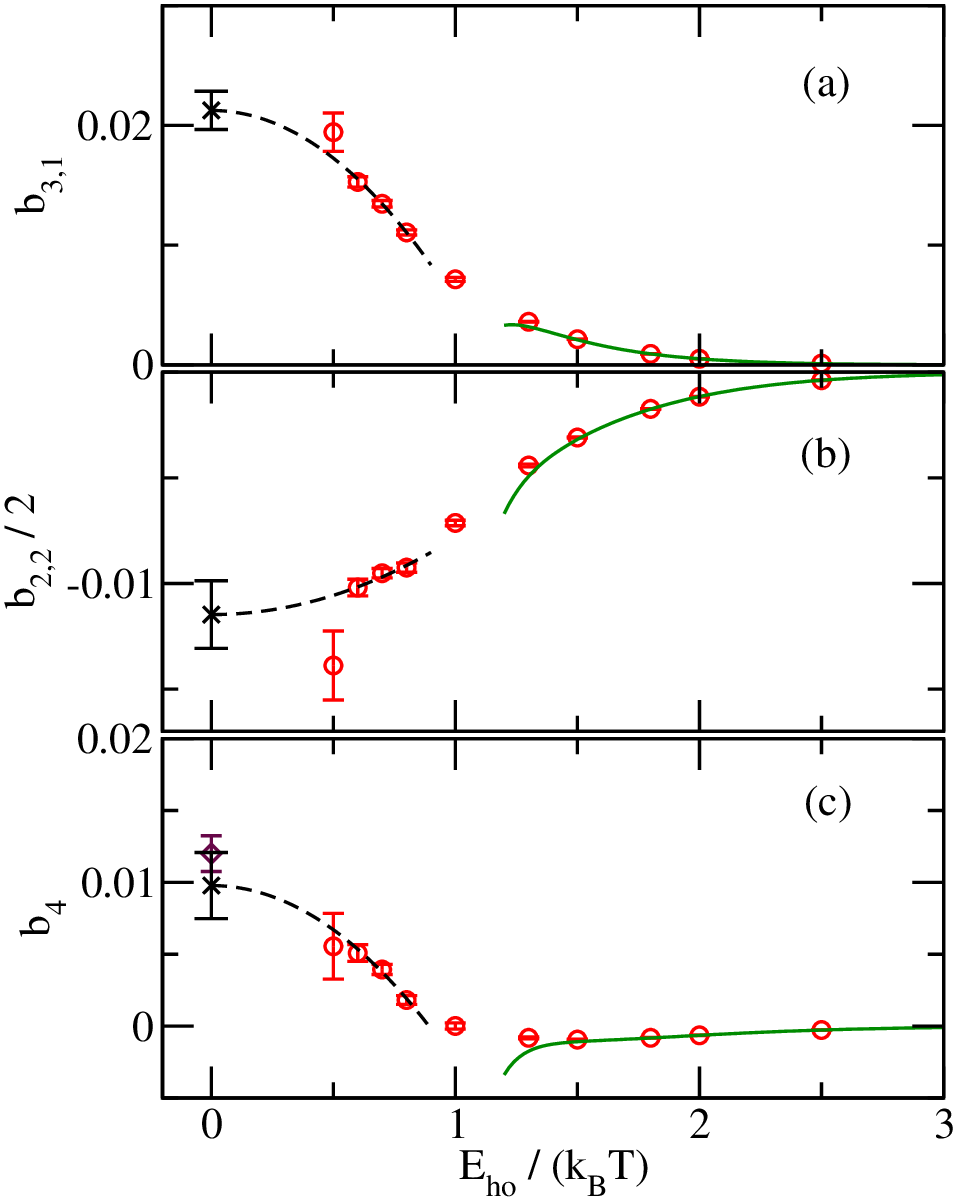}
\caption{(Color online)
  PIMC determination of the fourth-order virial coefficient. Circles in panels
  (a), (b), and (c) show $b_{3,1}$, $b_{2,2}/2$, and $b_4$, respectively,
  determined by our PIMC approach.
  The crosses in (a) and (b) show the $T\to\infty$ limit
  of the two-parameter
  fit (dashed line) to the PIMC data at the
  four highest temperatures. The dashed line and the cross
  in (c) show the sum of the fits from
  (a) and (b). The error bar in (c) is obtained 
  by error propagation.
  The diamond with error bar shows the experimental result from
  Ref.~\cite{mitexp}.
 }\label{figb4}
\end{figure}

We now discuss the determination of $b_4$.
The extrapolation of the raw data to the $\tau\to0$ limit is discussed in
the supplemental material~\cite{supnote}. Circles in Figs.~\ref{figb4}(a)
and~\ref{figb4}(b) show our PIMC results for $b_{3,1}$ and $b_{2,2}$,
respectively, as a function of the inverse temperature.
At low temperature, the PIMC results agree with the sum-over-states results
(solid lines), obtained using the data provided in Ref.~\cite{rakshit12}.
At all temperatures, $b_{3,1}$ is positive and $b_{2,2}$ is negative.
It has been shown that $b_{1,1}$ and $b_{2,1}$ are even functions 
of $E_{\text{ho}}/(k_BT)$~\cite{liu09,kevin12,b3bose}, 
and the conjecture of Ref.~\cite{endoarXiv15} implies that 
$b_{3,1}$ and $b_{2,2}$ are also even functions of 
$E_{\text{ho}}/(k_BT)$.
Thus, to obtain $b_{3,1}$ and $b_{2,2}$, we fit the data points
for the four highest temperatures to the form $a+b [E_{\text{ho}}/(k_BT)]^2$. The 
dashed lines in Figs.~\ref{figb4}(a) and \ref{figb4}(b) show 
the fits.
Since the data points at $k_BT=2E_{\text{ho}}$ have much larger error bars
than those at lower temperatures, the data points contribute comparatively 
little to the fit, which weighs each data point by the 
inverse of the square of its error
bar.
We find $b_{3,1}^0=0.0212(8)$ and $b_{2,2}^0/2=-0.0115(8)$, where the error bars reflect
the uncertainty of the fit.
We unfortunately do
not have sufficiently many data to include a $(k_B T)^{-4}$ term in
the fit.
Since the inclusion of a $(k_B T)^{-4}$ term in the fit could alter the
$T\to\infty$ result, we add an ad-hoc systematic error of 0.0008 to $b_{3,1}^0$
and $b_{2,2}^0/2$,
yielding $b_{3,1}^0=0.0212(16)$ and $b_{2,2}^0/2=-0.0115(16)$ [crosses in
Figs.~\ref{figb4}(a) and \ref{figb4}(b)].
To obtain $b_4$ [see Fig.~\ref{figb4}(c)], we combine $b_{3,1}$ and $b_{2,2}$.
Specifically, the circles and the fit are obtained by adding the data of
Figs.~\ref{figb4}(a) and \ref{figb4}(b) while the error bar of the cross at $T\to\infty$
is obtained
using standard error propagation.
$b_4$ displays an interesting temperature dependence: It is negative at
low temperature due to the dominance of $b_{2,2}$, vanishes at $k_BT\approx
E_{\text{ho}}$ due to a cancellation of $b_{3,1}$ and $b_{2,2}/2$, and is positive at
high temperature due to the dominance of $b_{3,1}$.
Our results resolve 
the discrepancy of the sign of $b_4$ 
between Ref.~\cite{rakshit12} and the experiments~\cite{ensexp,mitexp}.
Our extrapolated $b_4$ at infinite temperature is $b_{4}^0=0.0098(23)$,
which agrees with the experimental results of
$b_4^0=0.01203(125)$~\cite{mitexp} [diamond in Fig.~\ref{figb4}(c)] and
$b_4^0=0.01200(188)$~\cite{ensexp} (see
also Table~\ref{tab1}).
Using the LDA, we find $b_4^{\text{hom}}=0.078(18)$.

We now compare our results for $b_{3,1}^0$ and $b_{2,2}^0$ with the
literature. The diagrammatic approach~\cite{levinsen15} 
yields $b_{3,1}^0=0.025$,
which is within 2.5 standard deviations of our value, and $b_{2,2}^0/2=-0.018$,
which differs by a factor of about 1.5 (or many standard deviations)
from our value. This comparison suggests that the convergence of the
diagrammatic approach is slower for the (2,2) system than for the (3,1)
system.
The conjecture-based approach~\cite{endo15,endoarXiv15} yields
$b_{3,1}^0=0.02297(4)$, which agrees within error bars with
our value, and
$b_{2,2}/2=-0.0309(1)$, which differs by about a factor of 3 from our value.

\textit{Conclusion:}
This letter presented the PIMC determination of the fourth-order
virial coefficient of the trapped unitary two-component Fermi gas.
Our extrapolated infinite temperature result was found to agree
with experiments within error bars, which, to the best of our knowledge, is the
first numerical confirmation of the experimental determination of $b_4$.
The customized PIMC scheme, which allows for the treatment of Fermi gases with
zero-range interactions, can be applied to a variety of other situations.
Since
the zero-range density matrix can be constructed for arbitrary
$s$-wave scattering length $a_s$, 
the scheme can be used to study 
the finite-temperature characteristics of the BEC-BCS crossover of few-body Fermi gases.
Moreover, the algorithmic developments can be integrated into PIMC ground state calculations, providing a viable alternative to basis set expansion approaches.

{\em{Acknowledgement:}}
We are grateful to Kevin M. Daily for valuable correspondence, to Xiangyu Yin
for thoughtful comments on the manuscript, and to Yvan
Castin for correspondence related to Refs.~\cite{endoarXiv15} and
\cite{endoarXiv15v2}.
Support by the National
Science Foundation (NSF) through Grant No.
PHY-1415112
is gratefully acknowledged.
This work used the Extreme Science and Engineering
Discovery Environment (XSEDE), which is supported by
NSF Grant No. OCI-1053575, and the
WSU HPC.

{\em{Note added:}} After submission of this
paper, Endo and Castin revised their conjecture presented in Ref.~\cite{endoarXiv15}.
The new calculation yields $b_4^0=0.00775(10)$,
$b_{3,1}^0=0.02297(4)$, and $b_{2,2}^0/2=-0.0152(1)$~\cite{endoarXiv15v2}.


\begin{thebibliography}{50}
\bibitem [{inf()}]{infinitestdnote}%
  \BibitemOpen
  \href@noop {} {}\bibinfo {note} {For a simulation of finite length $l$, the
  standard deviation would not be following a Gaussian distribution, implying
  that the standard deviation would not decrease as $1/\sqrt{l}$ with
  increasing simulation length $l$.}\BibitemShut {Stop}%
\end{thebibliography}
%
\end{document}


\title{Supplemental Material \\
Path integral Monte Carlo determination of the fourth-order virial coefficient for unitary two-component 
  Fermi gas with zero-range interactions
}
\author{Yangqian Yan}
\affiliation{Department of Physics and Astronomy,
Washington State University,
  Pullman, Washington 99164-2814, USA}
\author{D. Blume}
\affiliation{Department of Physics and Astronomy, Washington State University, Pullman,
Washington 99164-2814, USA}

\date{\today }
\renewcommand{\theequation}{S\arabic{equation}}
\renewcommand{\thefigure}{S\arabic{figure}}
\renewcommand{\thetable}{S\arabic{table}}

\maketitle

\onecolumngrid
The notation employed in this Supplemental Material follows that introduced in
the main text.

\section{literature values of $b_4$}
Table~\ref{tabx}
\begin{table}[!pb]
  \centering
  \caption{
    Summary of literature results. The value reported in the respective
    reference is underlined. The conversion to other ``representations'' is
    done using Eqs.~(\ref{eqnoni})-(\ref{eqlda}).
    The column labeled ``Ref.'' refers to the bibliography of the main text.
}
\begin{tabular} 
  {c c c c c}
    \hline
    \hline
    \multicolumn{1}{c}{$b_4^{\text{hom}}$} & \multicolumn{1}{c}{$b_4^{\text{hom,tot}}$} & \multicolumn{1}{c}{$b_4^0$} & Ref. & comment\\
    \underline{0.096(15)}& 0.065(15) &  0.01200(188) & 19 & ENS experiment\\
     0.096(10)& \underline{0.065(10)} &  0.01203(125) & 26 & MIT experiment\\
     -0.016(4) & -0.04725(40) & \underline{-0.0020(5)} & 22 & sum-over-states approach\\
     \underline{0.06} & 0.02875& 0.0075 &23 & diagrammatic approach\\
     \underline{-0.063(1)} & -0.09425(10) & -0.007875(125) &24 & 3-body inspired
     conjecture\\
    \hline
    \hline
  \end{tabular}
  \label{tabx}
\end{table}
summarizes the literature results for the fourth-order virial
coefficient; this table is an extended version of Table I of the main text. The non-interacting contribution to the total fourth-order virial
coefficient $b_4^{\text{hom,tot}}$ of the homogeneous system is given by 
\begin{equation}
  b_n^{\text{hom},\text{ni}}=(-1)^{n+1}/n^{5/2};
  \label{eqnoni}
\end{equation}
the interacting part of the fourth-order virial coefficient $b_4^{\text{hom}}$
of the homogeneous system is defined through 
\begin{equation}
  b_n^{\text{hom}}=b_n^{\text{hom,tot}}-b_n^{\text{hom},\text{ni}}.
  \label{}
\end{equation}
The interacting part of the fourth-order virial coefficient $b_4^0$ of the
harmonically trapped system at high temperature and $b_4^{\text{hom}}$ are
related via (see also the main text),
\begin{equation}
  b_n^{\text{hom}}=n^{3/2} b_n^0.
  \label{eqlda}
\end{equation}

\section{Pair product approximation and zero-range density matrix}
Equation (9) of the main text writes the observable
$Q_{n_1,n_2}^{\text{ni}}/Q_{n_1,n_2}$ in terms of the density matrices 
$\rho^{\text{ni}}(\mathbf{R}_i,\mathbf{R}_{i+1};\tau)$ and $\rho(\mathbf{R}_i,\mathbf{R}_{i+1};\tau)$ of the non-interacting and unitary $(n_1,n_2)$-particle systems. To evaluate the density matrix by the PIMC approach, we use the pair product approximation~\cite{ceperleyrev},
\begin{eqnarray}
  \rho(\mathbf{R},\mathbf{R}';\tau)
  \approx
  \left(\prod_{j=1}^{n_1+n_2}\rho^{\text{sp}}(\mathbf{r}_j,\mathbf{r}'_j;\tau)
    \right)\times\nonumber \\
    \left(\prod_{j=1}^{n_1}\prod_{k=n_1+1}^{n_1+n_2}\bar{\rho}^{\text{rel}}(\mathbf{r}_{j}-\mathbf{r}_{k},\mathbf{r}'_{j}-\mathbf{r}'_{k};\tau)
    \right)
    ,
  \label{rhotot}
\end{eqnarray}
where $\rho^{\text{sp}}(\mathbf{r},\mathbf{r}';\tau)$ is the single-particle density matrix~\cite{ceperleyrev},
  \begin{eqnarray}
    \rho^{\text{sp}}(\mathbf{r},\mathbf{r}';\tau)=
    a_{\text{ho}}^{-3}\left[2\pi \sinh(\tau\hbar\omega)\right]^{-3/2}\times\nonumber \\
    \exp\left({-\frac{(\mathbf{r}^2+\mathbf{r}'^2)\cosh(\tau\hbar\omega)
    -2\mathbf{r}\cdot\mathbf{r}'}{2\sinh(\tau\hbar\omega)a_{\text{ho}}^2}}\right),
    \label{}
  \end{eqnarray}
  and $\bar{\rho}^{\text{rel}}(\mathbf{r},\mathbf{r}';\tau)$ is the reduced pair density matrix of the relative two-body problem with zero-range interaction~\cite{zr15},
\begin{equation}
    \bar{\rho}^{\text{rel}}(\mathbf{r},\mathbf{r}';\tau)= 1+\frac{2\hbar^2
    \tau}{m r r'}\exp\left(-\frac{m (r r'+\mathbf{r}\cdot \mathbf{r'})}{2\hbar^2 \tau}\right)
    .
  \label{zrpropagator}
\end{equation}
The density matrix $\rho^{\text{ni}}$ of the non-interacting system is given
by 
Eq.~(\ref{rhotot}) with $\bar{\rho}^{\text{rel}}$ replaced by 1.

\section{Extrapolation to the $\tau\to0$ limit and selected raw data}
As mentioned in the main text, to determine $b_n$ with comparable percentage
accuracy at all temperatures, $Q_{n_1,n_2}^{\text{ni}}/Q_{n_1,n_2}$ has to
be determined with increasing percentage accuracy with increasing temperature.
To  ensure that our results are free of systematic errors, the error
introduced by the $\tau\to0$ extrapolation has to be smaller than the error of
the extrapolation that arises from the statistical error of the individual
PIMC data points. To illustrate this, we consider the (2,1) system at the
highest temperature considered, i.e., at $k_B
T=2E_{\text{ho}}$.

\begin{figure}[!tp]
\includegraphics[angle=0,width=0.4\textwidth]{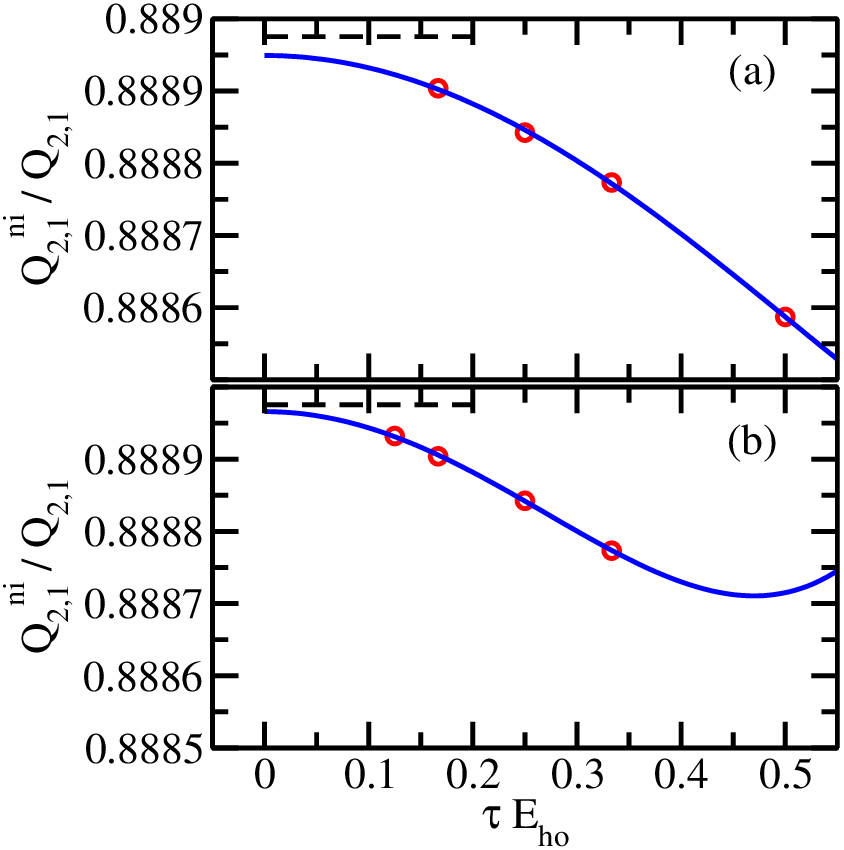}
\caption{(Color online)
  Benchmarking our PIMC results (circles) for the (2,1) system at unitarity
  through comparison with sum-over-states results.
  The observable $Q_{2,1}^{\text{ni}}/Q_{2,1}$
  as a function of the
  imaginary time step $\tau$ at temperature $k_BT=2E_{\text{ho}}$.
  The error bars (not shown) are smaller than the symbol size.
  In (a), the time steps correspond to $N=2$, $3$, $4,$ and $6$.
  In (b), the time steps correspond to $N=3$, $4$, $6,$ and $8$.
  The solid line shows the fourth-order polynomial fit of the form $a+b
  \tau^2+c\tau^4$.
  The dashed line shows the sum-over-states results.
 }\label{figerror}
\end{figure}

Circles in Fig.~\ref{figerror}(a) show $Q_{2,1}^{\text{ni}}/Q_{2,1}$, obtained by
our PIMC approach, as a function of the imaginary time step $\tau$ (the data
correspond to $N=2,3,4,$ and 6). The solid line shows a fourth-order fit of
the form $a+b \tau^2+c \tau^4$ to our PIMC data.
The extrapolated $\tau\to0$ value of 0.888949(8), where the error bar accounts
for the statistical uncertainty of the PIMC data, deviates by about 3 standard
deviations (or $0.003\%$) from the sum-over-states result of 0.8889755. We attribute the
discrepancy to the fact that the $\tau$ considered are not small enough for
the fourth-order fit to be fully reliable. To corroborate this interpretation,
we (i) employ a sixth-order fit and (ii) apply the fourth-order fit to PIMC
data for smaller $\tau$. The sixth-order fit (using, as before, the data
corresponding to $N=2,3,4,$ and 6) yields 0.888964(19), in agreement with the
sum-over-states result.
Note, however, that the error bar is much larger than that resulting from the
fourth-order fit; the reason is that we are attempting to determine four fit
parameters using just four 
data points. Performing a fourth-order fit to the PIMC
data for $N=3,4,6$, and $8$ yields 0.888966(8), which almost agrees with the
sum-over-states approach within error bar and with an error bar that is comparable to
our previous fourth-order
fit [see Fig.~\ref{figerror}(b)].
This analysis suggests that our PIMC calculations are free of systematic
errors provided we go to sufficiently small $\tau$.

Table~\ref{tab1} lists the PIMC raw data for the (3,1) and (2,2) systems at
\begin{table}[!tbp]
  \centering
  \caption{
Selected PIMC raw data.
Columns 1 and 2 show the inverse temperature $E_{\text{ho}}/(k_B T)$ and 
the number of 
imaginary time slices $N$, respectively.
Columns 3 and 4 show the observables $Q_{3,1}^{\text{ni}}/Q_{3,1}$ and
$Q_{2,2}^{\text{ni}}/Q_{2,2}$
for the (3,1) and (2,2) systems,
respectively.
}
\begin{tabular} 
{c c c c}
    \hline
    \hline
    $E_{\text{ho}}/(k_B T)$ & $N$ & $Q_{3,1}^{\text{ni}}/Q_{3,1}$ & $Q_{2,2}^{\text{ni}}/Q_{2,2}$ \\
    0.5   & 2 &   0.8413081(35)&0.7940517(46)  \\
    0.5   & 3 &   0.8418155(43)&0.7946990(43)  \\
    0.5   & 4 &   0.8420157(41)&0.7949482(45)  \\
    0.5   & 6 &   0.8421806(41)&0.7951732(53)  \\
    \hline                                    
    0.6   & 3 &   0.754475(16) &0.686274(11)  \\
    0.6   & 4 &   0.754955(12) &0.686860(11)  \\
    0.6   & 5 &   0.755218(12) &0.687174(14)  \\
    0.6   & 7 &   0.755450(13) &0.687445(14)  \\
    0.6   & 9 &   0.755591(15) &0.687587(14)  \\
    \hline                                    
    0.7   & 3 &   0.658547(24) &0.571429(24)  \\
    0.7   & 4 &   0.659464(22) &0.572473(14)  \\
    0.7   & 6 &   0.660203(26) &0.573329(22)  \\
    0.7   & 9 &   0.660583(29) &0.573764(23)  \\
    \hline                                    
    0.8   & 4 &   0.563935(34) &0.462752(36)  \\
    0.8   & 5 &   0.564662(35) &0.463530(33)  \\
    0.8   & 7 &   0.565379(36) &0.464433(32)  \\
    0.8   & 9 &   0.565708(38) &0.464757(33)  \\
    \hline
    \hline
  \end{tabular}
  \label{tab1}
\end{table}
various temperatures (the data for low temperatures are not shown). We report
the observables $Q_{3,1}^{\text{ni}}/Q_{3,1}$ and $Q_{2,2}^{\text{ni}}/Q_{2,2}$
for various time slices. 
For $E_{\text{ho}}/(k_B T)=0.6$, $0.7$, and $0.8$, 
the largest number of time slices considered is $N_{\text{max}}=9$.
For $E_{\text{ho}}/(k_B T)=0.5$, our available
computing resources limit us to $N_{\text{max}}=6$, resulting 
in reduced accuracy of the observables.
\begin{table}[!tbp]
  \centering
  \caption{
Selected extrapolated PIMC results.
Columns 1 and 2 show the inverse
temperature $E_{\text{ho}}/(k_B T)$ and the order used in the extrapolation, respectively.
Columns 3 and 5 show the extrapolated $\tau\to0$ observables 
$Q_{3,1}^{\text{ni}}/Q_{3,1}$ and $Q_{2,2}^{\text{ni}}/Q_{2,2}$
for the  (3,1) and (2,2)
systems, respectively.
Columns 4 and 6 show the resulting subcluster contributions $b_{3,1}$ and $b_{2,2}/2$, respectively,
to the fourth-order virial coefficient.
}
\begin{tabular} 
{c c c c c c}
    \hline
    \hline
     $E_{\text{ho}}/(k_B T)$ &order &   $Q_{3,1}^{\text{ni}}/Q_{3,1}$ & $b_{3,1}$  &$Q_{2,2}^{\text{ni}}/Q_{2,2}$ &$b_{2,2}/2$ \\
    0.5 &6& 0.842330(15) & 0.0194(16) &0.795393(18) & $-0.0139(16)$ \\
    0.6 &4& 0.755751(18) & 0.0153(4)  &0.687775(18) & $-0.0102(4)$  \\
    0.7 &4& 0.660877(39) & 0.0135(3)  &0.574108(30) & $-0.0095(2)$  \\
    0.8 &4& 0.566227(82) & 0.0111(2)  &0.465415(73) & $-0.0093(2)$  \\
    \hline
    \hline
  \end{tabular}
  \label{tab2}
\end{table}

For $E_{\text{ho}}/(k_B T)=0.6$, $0.7$, and $0.8$, we perform fourth-order
fits to the $\tau$-dependent $Q_{3,1}^{\text{ni}}/Q_{3,1}$ and 
$Q_{2,2}^{\text{ni}}/Q_{2,2}$ data listed in Table~\ref{tab1},
yielding extrapolated $\tau\to0$ values with error bars between 0.0024\% and
0.016\%.
We estimate, based on our tests for the three-body system, that these
statistical errors are larger than the systematic error, which arises from the
use of the fourth-order fit. Hence the systematic uncertainty can be
neglected.
For $E_{\text{ho}}/(k_B T)=0.5$, a fourth-order fit to the data given in
Table~\ref{tab1} yields error bars of 0.0008\% and 0.001\% for $Q_{3,1}^{\text{ni}}/Q_{3,1}$ and 
$Q_{2,2}^{\text{ni}}/Q_{2,2}$, respectively.
Since we estimate the systematic fit uncertainty to be, based on our analysis
for the (2,1) system, about $0.003\%$, we deem the fourth-order fit
unreliable.
Using a sixth-order fit (which yields a larger error bar), we find the values
listed in Table~\ref{tab2}.

%